\documentclass{article}
\usepackage{mirex2010,cite}
\usepackage{graphicx}
\usepackage{amsmath}
\usepackage{amsfonts}
\usepackage{amssymb}
\usepackage{graphicx}
\usepackage{url}
\usepackage{multirow}
\usepackage{multicol}
\title{Meta-song evaluation for chord recognition}

  {Yizhao Ni$^1$, Matt Mcvicar$^1$, Raul Santos-Rodriguez$^2$ and Tijl De Bie$^1$} {1. Intelligent Systems Lab\\ Department of Engineering Mathematics \\ University of Bristol, U. K.}
  {2. Signal Theory and Communications Department\\ Universidad Carlos III de Madrid\\ Spain}

\begin{document}
\maketitle
\begin{abstract}
We present a new approach to evaluate chord recognition systems on songs which do not have full annotations. The principle is to use online chord databases to generate high accurate ``pseudo annotations" for these songs and compute ``pseudo accuracies" of test systems. Statistical models that model the relationship between ``pseudo accuracy" and real performance are then applied to estimate test systems' performance. The approach goes beyond the existing evaluation metrics, allowing us to carry out extensive analysis on chord recognition systems, such as their generalizations to different genres. In the experiments we applied this method to evaluate three state-of-the-art chord recognition systems, of which the results verified its reliability.
\end{abstract}

\section{Introduction}
In recent years, audio chord recognition has become a very active field \cite{labrosa_mirex_2010,MM_thesis,probability_framework_key_chord_prediction_2007,unified_key_chord_prediction_2007,local_key_estimation_2009,rocher_key_estimation_2010} due to the increasing popularity of Music Information Retrieval (MIR) with applications using mid-level tonal features has established chord recognition as a useful and challenging task.

Generally speaking, chord recognition is a task of automatically detecting chord labels and boundaries from the audio of a musical piece. The process involves segmenting a song into a high time resolution sequence of windows (known as \textit{frames}), after which machine learning techniques (e.g.~Hidden Markov Models) are utilized to detect chord label for each frame, based on the features extracted and the local context. The chord predictions can then be evaluated via frame-wise accuracies, if the ground truth annotation of the song is available.

The annual MIREX (Music Information Retrieval Evaluation eXchange)   competition\footnote{\url{http://www.music-ir.org/mirex/wiki/MIREX_HOME}} has a task dedicated
to chord recognition, where participants attempt to generate chord predictions for a collection of songs. In the most recent competitions, the dataset used is a collection of Beatles, Queen and Zweieck songs, of which the \textit{ground truth annotations} are available. Due to the limited amount of data, existing chord recognition systems (referred as \textit{test systems }in the paper) are usually trained and tested on the same songs, inevitably causing
over-fitting on this dataset. Meanwhile, the evaluation is also heavily constrained by the simplicity of the data. For example, most of the songs in the dataset are from Rock genre, implying that the performance lacks generalization to other genres.

To resolve these problems, the simplest, but most costly and least scalable solution would be to obtain more fully annotated data, paying trained musicians to annotate new songs. Alternatively, we propose using a methodologically more challenging but cheaper and scalable approach: \textit{meta-song evaluation}, which makes use of large and freely available online chord databases, such as \textit{E-chords}\footnote{\url{http://www.e-chords.com/}} to help evaluate test systems. The principle is to automatically generate chord annotations for new songs of which the chord sequences are available on these databases. The songs and the generated annotations are then used to estimate test systems' performance via statistical theories.

However, chord sequences from such databases are generally less directly usable than those produced by musicians, since exact timings of chords are absent and sometimes the chord sequences are affected by various types of errors and omissions. Hence, a system, referred to as \textit{reference system} in the paper, is required to generate high accurate annotations from these untimed chord sequences. As demonstrated in our previous work \cite{JA_journal}, we have designed a variety of reference systems from which the generated annotations are more accurate than most of the existing chord recognition systems. We regard these high  accurate, but not perfect annotations as \textit{pseudo annotations}. In the rest of this paper, we will show how to make use of these pseudo annotations to comprehensively evaluate performances of different test systems.

\begin{figure}
 \centerline{
 \includegraphics[scale=0.33]{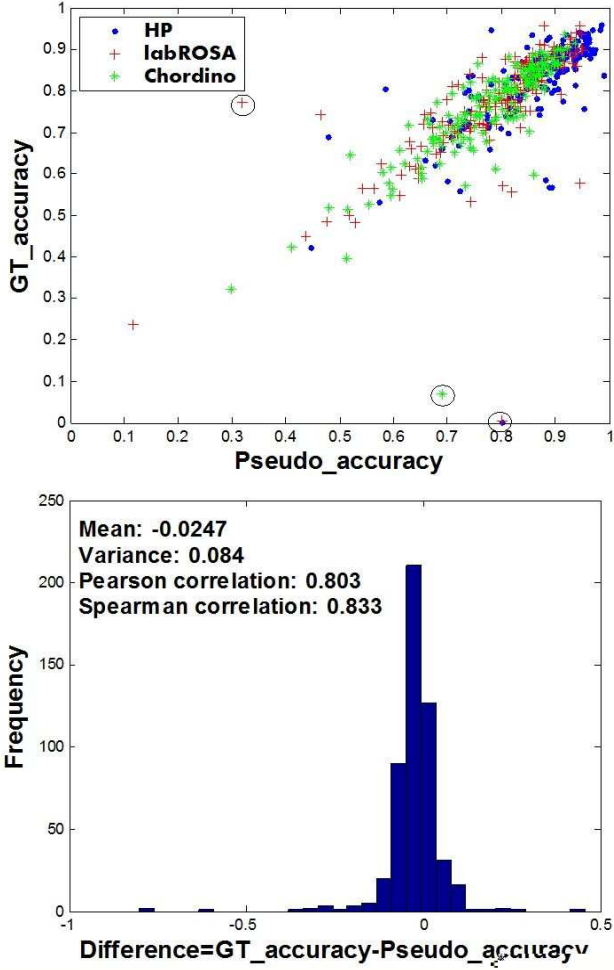}}
 \caption{The relationship between the pseudo accuracies and the real GT accuracies on $175$ The Beatles songs. The three systems are: A. HP, B. labROSA and C. Chordino. The pseudo annotations of the songs are generated by the Jump Alignment method \cite{JA_journal}, which has been shown to produce more accurate chord predictions than all other systems. Note that there are some outliers (represented by circled points on the top figure), of which the online chord sequences are less informative (e.g.~the chord sequence only records the solo of the song). In these cases, the resulting pseudo accuracies are not well-correlated with the GT accuracies. How to reduce and eliminate these outliers will be investigated in our future work. }
 \label{fig:PACC_vs_GTACC}
\end{figure}

\section{Mathematical framework}
We use $y^{A}_{i}$ and $x^{A}_{i}$ to denote the \textit{ground truth (GT) accuracy} and the \textit{pseudo accuracy} (i.e.~the accuracy of system's prediction compared to pseudo annotation) of system $A$'s chord prediction for the $i$-th song. Then for each system we obtain two sets of data:  a validation set $\{x^{A}_{i},y^{A}_{i}\}_{i=1}^{n}$ and a test set $\{x^{A}_{j}\}_{j=n+1}^{n+m}$. Note that we only have ground truth annotations on the validation set such that generally $m\gg n$. The test system pool is denoted by $A\in\mathcal{A}$.

One observation from the validation set is that the pseudo accuracies are highly correlated with the GT accuracies (see Figure \ref{fig:PACC_vs_GTACC}), as long as the pseudo annotations are accurate enough.  In the ideal case, if all pseudo annotations are $100\%$ accurate, the pseudo accuracies will converge to the GT accuracies. Inspired by this observation, we propose three mathematical frameworks to model the relationship between GT and pseudo accuracies on the validation set, which can then be applied to estimate GT accuracies on the test set.

\subsection{Single Gaussian model}\label{sec:single_gaussian_model}
This model assumes that the pairs $(x_{i},y_{i})$ generated by all systems $\mathcal{A}$ are sampled \textit{i.i.d} from a single Gaussian distribution $\mathcal{N}(\mu,\sigma^{2})$. That is
\begin{equation}
y^{A}_{i}= x^{A}_{i}+\mu+\epsilon_{i},\textrm{ }1\leq i \leq n,\textrm{ }\epsilon_{i}\sim\mathcal{N}(0,\sigma^{2}),\textrm{ }\forall A\in\mathcal{A}.
\end{equation}
The parameters of the distribution $\mu$ can then be easily estimated by \textit{least square} \cite{applied_linear_statistical_models}, resulting in
$$
\bar{\mu}=\frac{1}{|\mathcal{A}|n}\sum\limits_{A\in\mathcal{A}}\sum\limits_{i=1}^{n}(y^{A}_{i}-x^{A}_{i}).
$$
The unbiased estimator of $\sigma^{2}$ can be calculated by the formula
$$
\bar{\sigma}^{2}=\frac{1}{|\mathcal{A}|n-1}\sum\limits_{A\in\mathcal{A}}\sum\limits_{i=1}^{n}(y^{A}_{i}-x^{A}_{i}-\bar{\mu})^{2}.
$$

Using the parameters $(\bar{\mu},\bar{\sigma}^{2})$ estimated from the validation set, we are able to predict the GT accuracies $\{y^{A}_{j}\}_{j=n+1}^{n+m}$ on the test set using the linear regression theory. Let $\bar{x}=\frac{1}{|\mathcal{A}|n}\sum\limits_{A\in\mathcal{A}}\sum\limits_{i=1}^{n}x^{A}_{i}$ and $s^{2}_{x}=\frac{1}{|\mathcal{A}|n-1}\sum\limits_{A\in\mathcal{A}}\sum\limits_{i=1}^{n}(x^{A}_{i}-\bar{x})^2$, the following Gaussian distribution holds for all test examples\footnote{From a purely probabilistic perspective, the test examples follow a student-t distribution instead of a Gaussian. But since $n$ is large enough ($n>100$) in our case, we approximate the student-t distribution as a Gaussian.} \cite{applied_linear_statistical_models}:
\begin{equation}\label{eq:S_gaussian}
y^{A}_{j}-x^{A}_{j}-\bar{\mu}\sim \mathcal{N}\Big(0,\bar{\sigma}^{2}\big(1+\frac{1}{|\mathcal{A}|n}+\frac{(x^{A}_{j}-\bar{x})^2}{(|\mathcal{A}|n-1)s^2_{x}}\big)\Big).
\end{equation}
Therefore, with probability $1-\alpha$ the confidence interval of $y^{A}_{j}$ is
\begin{equation}
y^{A}_{j}=x^{A}_{j}+\bar{\mu}\pm Q(1-\alpha)\bar{\sigma}\left(1+\frac{1}{|\mathcal{A}|n}+\frac{(x^{A}_{j}-\bar{x})^2}{(|\mathcal{A}|n-1)s^2_{x}}\right)^{1/2},
\end{equation}
where $Q$ denotes a normal quantile function $Q(p)=\inf\{y\in \mathbb{R}:p\leq Pr(Y\leq y)\}$.

We then extend Eq. (\ref{eq:S_gaussian}) and estimate the mean accuracy of the test set $\bar{y}^{A}=\frac{1}{m}\sum\limits_{j=n+1}^{n+m}y^{A}_{j}$ using
$$
\bar{y}^{A}-\bar{x}^{A}-\bar{\mu}\sim \mathcal{N}(0,\hat{\sigma}^{2}_{A}),
$$
with $\bar{x}^{A}=\frac{1}{m}\sum\limits_{j=n+1}^{n+m}x^{A}_{j}$ and $\hat{\sigma}^{2}_{A}=\bar{\sigma}^{2}\frac{\sum\limits_{j=n+1}^{n+m}1+\frac{1}{|\mathcal{A}|n}+\frac{(x^{A}_{j}-\bar{x})^2}{(|\mathcal{A}|n-1)s^2_{x}}}{m^2}$.
Again with probability $1-\alpha$ the confidence interval of $\bar{y}^{A}$ is
\begin{equation}\label{eq:S_mean_acc_estimation}
\bar{y}^{A} = \bar{x}^{A}+\bar{\mu}\pm Q(1-\alpha)\hat{\sigma}_{A}.
\end{equation}

Apart from estimating the confidence interval of the GT accuracies, the Gaussian distribution also allows us to compare two systems $A$ and $B$, by means of estimating the confidence interval of $\bar{y}^{A}-\bar{y}^{B}$ using
$$\bar{y}^{A}-\bar{y}^{B}\sim \mathcal{N}(\bar{x}^{A}-\bar{x}^{B},\hat{\sigma}^{2}_{A}+\hat{\sigma}^{2}_{B}).$$
This yields the following confidence interval with probability $1-\alpha$
\begin{equation}\label{eq:S_acc_comparison}
\bar{y}^{A}-\bar{y}^{B}=\bar{x}^{A}-\bar{x}^{B}\pm Q(1-\alpha)\sqrt{\hat{\sigma}^{2}_{A}+\hat{\sigma}^{2}_{B}}.
\end{equation}

The advantage of the single Gaussian model is that it makes use of $|\mathcal{A}|$ times data to estimate the Gaussian parameters, which is expected to provide more robust estimation. However, as we observed in the experiments, test systems that are closer to the reference system generally got higher pseudo accuracies than the others. In this case, the GT accuracies estimated by the single Gaussian model would bias towards these systems.

\subsection{Individual Gaussian model}\label{sec:individual_gaussian_model}
To reduce or eliminate the effect of such biases, we proposed a variant of the single Gaussian model, fitting individual Gaussians to different test systems. Mathematically, the GT accuracy $y^{A}_{i}$ is now modelled as
\begin{equation}
y^{A}_{i}= x^{A}_{i}+\mu_{A}+\epsilon_{i},\textrm{ }1\leq i \leq n,\textrm{ }\epsilon_{i}\sim\mathcal{N}(0,\sigma_{A}^{2}),\textrm{ }\forall A\in\mathcal{A},
\end{equation}
where the parameters $(\mu_{A},\sigma_{A}^{2})$ can be learnt from the validation data $\{x_{i}^{A},y_{i}^{A}\}_{i=1}^{n}$:
\begin{equation}
\left\{
\begin{array}{l}
\bar{\mu}_{A}=\frac{1}{n}\sum\limits_{i=1}^{n}(y^{A}_{i}-x^{A}_{i})\\
\bar{\sigma}_{A}^{2}=\frac{1}{n-1}\sum\limits_{i=1}^{n}(y^{A}_{i}-x^{A}_{i}-\bar{\mu})^{2}\\
\end{array}\right., \forall A\in\mathcal{A}.
\end{equation}

Here we denote $\bar{x}=\frac{1}{n}\sum\limits_{i=1}^{n}x^{A}_{i}$ and $s^{2}_{x}=\frac{1}{n-1}\sum\limits_{i=1}^{n}(x^{A}_{i}-\bar{x})^2$, then following the same procedure as described in Section \ref{sec:single_gaussian_model} we obtain
$$
y^{A}_{j}-x^{A}_{j}-\bar{\mu}_A\sim \mathcal{N}\Big(0,\bar{\sigma}_{A}^{2}\big(1+\frac{1}{n}+\frac{(x^{A}_{j}-\bar{x})^2}{(n-1)s^2_{x}}\big)\Big),
$$
and with probability $1-\alpha$ the confidence interval of $y^{A}_{j}$ is
$$y^{A}_{j}=x^{A}_{j}+\bar{\mu}_{A}\pm Q(1-\alpha)\bar{\sigma}_{A}\big(1+\frac{1}{n}+\frac{(x^{A}_{j}-\bar{x})^2}{(n-1)s^2_{x}}\big)^{1/2}.
$$

Similarly, we have
\begin{equation}\label{eq:I_mean_acc_estimation}
\bar{y}^{A} = \bar{x}^{A}+\bar{\mu}\pm Q(1-\alpha)\hat{\sigma}_{A},
\end{equation}
with $\bar{x}^{A}=\frac{1}{m}\sum\limits_{j=n+1}^{n+m}x^{A}_{j}$ and $\hat{\sigma}^{2}_{A}=\bar{\sigma}_{A}^{2}\frac{\sum\limits_{j=n+1}^{n+m}1+\frac{1}{n}+\frac{(x^{A}_{j}-\bar{x})^2}{(n-1)s^2_{x}}}{m^2}$.

To compare the two systems $A$ and $B$, we now have a term $\bar{\mu}_{A}-\bar{\mu}_{B}$ to reduce the effect of the biases, yielding
$$\bar{y}^{A}-\bar{y}^{B}\sim \mathcal{N}(\bar{x}^{A}-\bar{x}^{B}+\bar{\mu}_{A}-\bar{\mu}_{B},\hat{\sigma}^{2}_{A}+\hat{\sigma}^{2}_{B}).$$
We derive the following confidence interval with probability $1-\alpha$
\begin{equation}\label{eq:I_acc_comparison}
\bar{y}^{A}-\bar{y}^{B}=\bar{x}^{A}-\bar{x}^{B}+\bar{\mu}_{A}-\bar{\mu}_{B}\pm Q(1-\alpha)\sqrt{\hat{\sigma}^{2}_{A}+\hat{\sigma}^{2}_{B}}.
\end{equation}

\subsection{Linear regression model}\label{sec:linear_regression}
Apart from applying different $\mu_{A}$ to eliminate the biases, one can also learn the slope of the regression line to better fit the validation samples. Mathematically, the relationship between $y_{i}^{A}$ and $x_{i}^{A}$ is now formulated as
\begin{equation}
y^{A}_{i}= a_{A}x^{A}_{i}+b_{A}+\epsilon_{i},\textrm{ }1\leq i \leq n,\textrm{ }\epsilon_{i}\sim\mathcal{N}(0,\sigma_{A}^{2}),\textrm{ }\forall A\in\mathcal{A},
\end{equation}
where the parameters $(a_{A},b_{A},\sigma_{A}^2)$ can be estimated by least square: $\forall A\in\mathcal{A}$
\begin{equation}
\left\{
\begin{array}{l}
\bar{a}_{A}=\frac{\sum\limits_{i=1}^{n}(y^{A}_{i}-\bar{y})(x^{A}_{i}-\bar{x})}{\sum\limits_{i=1}^{n}(x^{A}_{i}-\bar{x})^2}\\
\bar{b}_{A}=\bar{y}-\bar{a}_{A}\bar{x}\\
\bar{\sigma}_{A}^{2}=\frac{1}{n-2}\sum\limits_{i=1}^{n}(y_{i}^{A}-\bar{a}_{A}x_{i}^{A}-\bar{b}_{A})^2\\
\end{array}\right.,
\end{equation}
with $\bar{x}=\frac{1}{n}\sum\limits_{i=1}^{n}x_{i}^{A}$ and $\bar{y}=\frac{1}{n}\sum\limits_{i=1}^{n}y_{i}^{A}$.

Given the parameters, a test sample $y_{j}^A$ follows the following Gaussian distribution
$$
y^{A}_{j}-\bar{a}_{A}x^{A}_{j}-\bar{b}_A\sim \mathcal{N}\Big(0,\bar{\sigma}_{A}^{2}\big(1+\frac{1}{n}+\frac{(x^{A}_{j}-\bar{x})^2}{(n-1)s^2_{x}}\big)\Big),
$$
and its confidence interval is of the form
$$y^{A}_{j}=\bar{a}_{A}x^{A}_{j}+\bar{b}_A\pm Q(1-\alpha)\bar{\sigma}_{A}\big(1+\frac{1}{n}+\frac{(x^{A}_{j}-\bar{x})^2}{(n-1)s^2_{x}}\big)^{1/2}.
$$

Analogously to that presented in Section \ref{sec:individual_gaussian_model}, the mean accuracy $\bar{y}^{A}$ satisfies
$
\bar{y}^{A}-\bar{a}_{A}\bar{x}^{A}-\bar{b}_{A}\sim \mathcal{N}(0,\hat{\sigma}_{A}^{2}),
$
and with probability $1-\alpha$ the confidence interval of $\bar{y}^{A}$ is
\begin{equation}\label{eq:L_mean_acc_estimation}
\bar{y}^{A} = \bar{a}_{A}\bar{x}^{A}+\bar{b}_{A}\pm Q(1-\alpha)\hat{\sigma}_{A}.
\end{equation}

To compare the two systems $A$ and $B$, the confidence interval of $\bar{y}^{A}-\bar{y}^{B}$ is now calculated by
\begin{equation}\label{eq:L_acc_comparison}
\bar{y}^{A}-\bar{y}^{B}=\bar{a}_{A}\bar{x}^{A}-\bar{a}_{B}\bar{x}^{B}+\bar{b}_{A}-\bar{b}_{B}\pm Q(1-\alpha)\sqrt{\hat{\sigma}^{2}_{A}+\hat{\sigma}^{2}_{B}}.
\end{equation}

\begin{table*}[htb]
\begin{center}
\begin{tabular}{|c|c|c|c|c|}
\hline
System&S model [$\%$]& I model [$\%$]&L model [$\%$]& GT acc [$\%$]\\
\hline
HP & $82.6\pm1.3$&$82.2\pm1.3$ &$82.2\pm1.3$&$82.2$\\
labROSA &$76.6\pm1.3$&$77.6\pm1.4$&$77.6\pm1.3$&$77.6$\\
Chordino & $75.2\pm1.3$ &$76.2\pm1.0$ &$76.2\pm1.0$ &$76.2$\\
Consensus &$82.3\pm1.3$&$82.7\pm1.2$&$82.7\pm1.2$&$82.7$\\
\hline
\end{tabular}

\vspace{0.1cm}

\begin{tabular}{|c|c|c|c|c|}
\hline
System&S model [$\%$]& I model [$\%$]&L model [$\%$]& GT acc [$\%$]\\
\hline
HP - labROSA & $6.6\pm1.9$&$4.6\pm1.9$ &$4.6\pm1.9$&$4.6$\\
HP - Chordino &$8.0\pm1.7$&$6.0\pm1.6$&$6.0\pm1.6$&$6.0$\\
HP - Consensus&$0.3\pm1.8$&$-0.5\pm1.8$&$-0.5\pm1.8$&$-0.5$\\
labROSA - Chordino & $1.3\pm0.9$ &$1.4\pm1.7$ &$1.4\pm1.6$ &$1.4$\\
\hline
\end{tabular}
\end{center} \caption{Upper table: the estimation of performance of HP, labROSA and Chordino on the validation set. The first three columns are the estimated mean GT accuracies using Eq. (\ref{eq:S_mean_acc_estimation}), (\ref{eq:I_mean_acc_estimation}) and (\ref{eq:L_mean_acc_estimation}) respectively, where the confidence level is fixed at $95\%$. The forth column is the real GT accuracies. Lower table: the comparison of performances between test systems, using Eq. (\ref{eq:S_acc_comparison}), (\ref{eq:I_acc_comparison}), (\ref{eq:L_acc_comparison}) and real GT accuracy differences respectively. }\label{tab:verification}
\end{table*}

\section{Experiments}
Here we summarize the main experiments conducted, which consist of the estimation and comparison of the performances of three pre-trained chord recognition systems: the HP system \cite{HP_system_arxiv} that is trained on the audio dataset used  in the MIREX Chord Detection task 2010\footnote{\url{http://www.music-ir.org/mirex/wiki/2010:Audio_Chord_Estimation}}, the labROSA system \cite{labrosa_mirex_2010} which is trained on the same dataset, and finally Chordino, a freely-available pre-trained chord recognition system \cite{nnls}. The reference system used to generate pseudo annotations is the Jump Alignment (JA) method \cite{JA_journal}, which has shown to produce more accurate chord predictions than all other systems, by means of using the online chord database E-chords. The validation set consists of $175$ The Beatles' songs, of which we have both ground truth and pseudo annotations. This set is used to learn the parameters of the single Gaussian (S), the individual Gaussian (I) and the linear regression (L) models. The test set consists of $1840$ songs from a variety of genres, of which we can only derive pseudo annotations using JA. The objective of the experiments is to estimate and compare the GT accuracies of the three systems on the test set, in terms of the S, I and L models.

\subsection{Verification}
We first regarded the validation set as test set so as to verify the confidence intervals estimated by S, I and L models. The results are shown in Table \ref{tab:verification}. All real GT accuracies fall in the estimated interval with a $95\%$ confidence level, verifying the reliability of the models. We also observed from Table \ref{tab:verification} (lower table) that S model biases towards HP as expected, because HP shared the same chromagram features with the JA method. This bias was then removed by using I/L models.

\subsection{Performance estimation}
\begin{figure}
 \centerline{
 \includegraphics[scale=0.3]{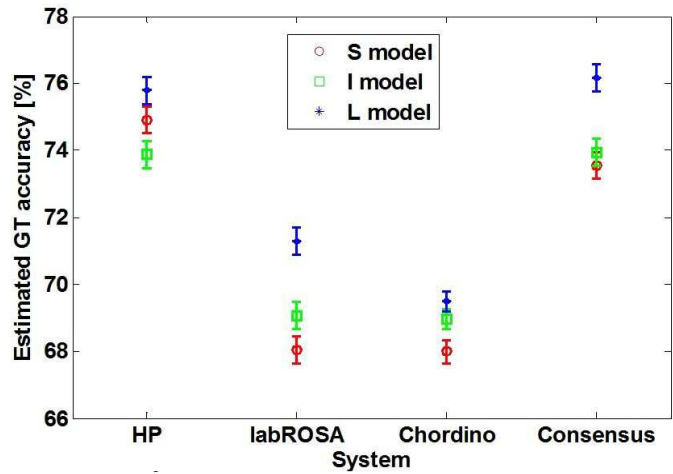}}
 \caption{The estimated mean GT accuracies of the test systems on a large test set ($1840$ songs).}
 \label{fig:estimated_GTacc_on_musicDB}
\end{figure}

We then estimated and compared performances of the systems on the large test set, of which the ground truth annotations are not available. Again, we first estimated the mean GT accuracies of the three systems, in terms of S/I/L models and pseudo annotations generated by JA. The results are illustrated in Figure \ref{fig:estimated_GTacc_on_musicDB}. We observed that the estimated accuracies between labROSA and Chordino are highly overlapped, indicating a similar performance of the two systems. Alternatively, there is a large gap between HP and the other two systems, implying the superiority of the HP system. We also observed that S model ranked higher than I model on HP, which is different from the cases for the other systems. This implies a bias towards HP, which however, was eliminated by I/L models.

\begin{figure*}
 \centerline{
 \includegraphics[scale=0.7]{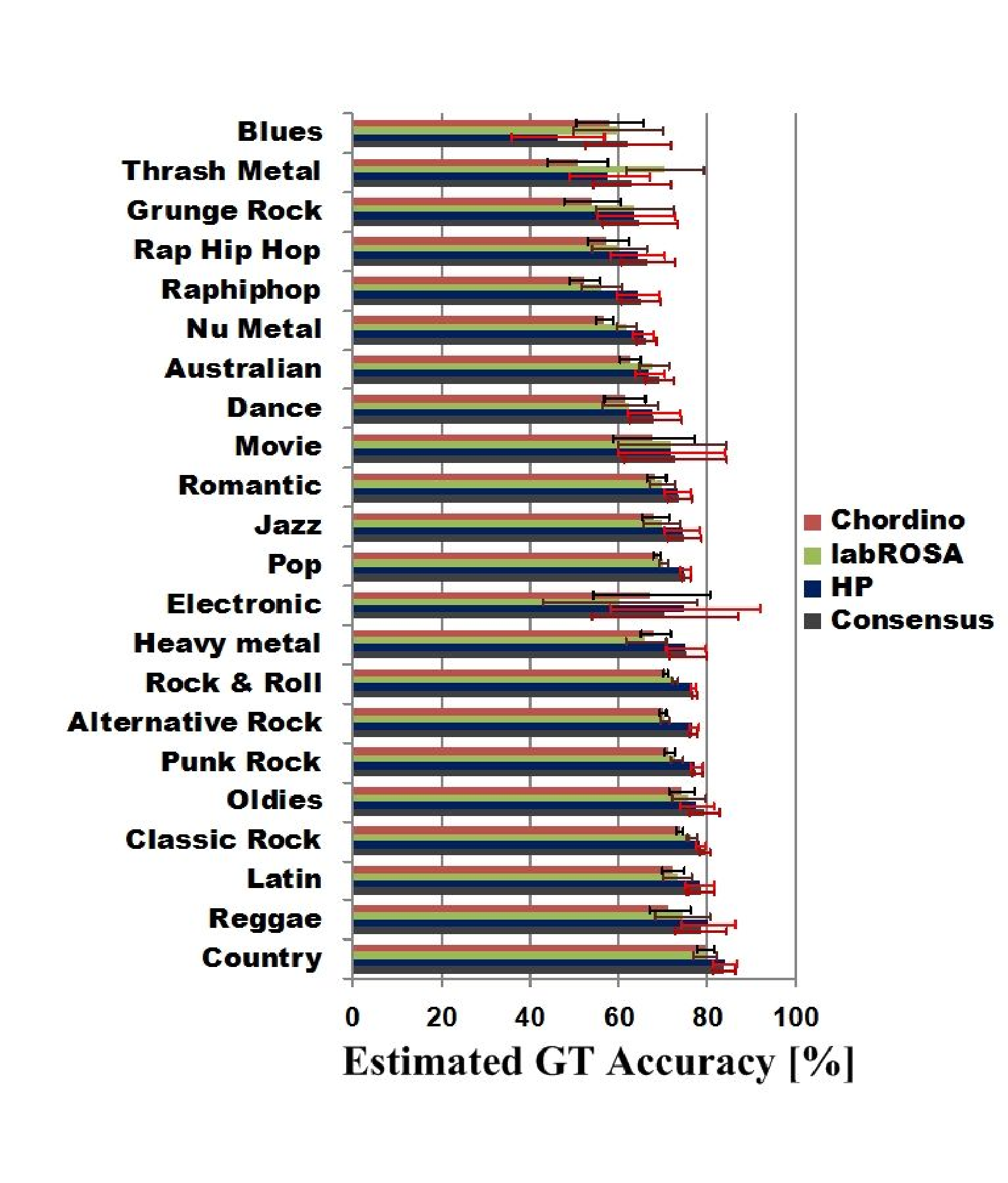}}
 \caption{Estimated GT accuracies of the test systems on each genre, using S model. }
 \label{fig:genre_performance_musicDB_singleGaussian}
\end{figure*}

\begin{figure*}
 \centerline{
 \includegraphics[scale=0.7]{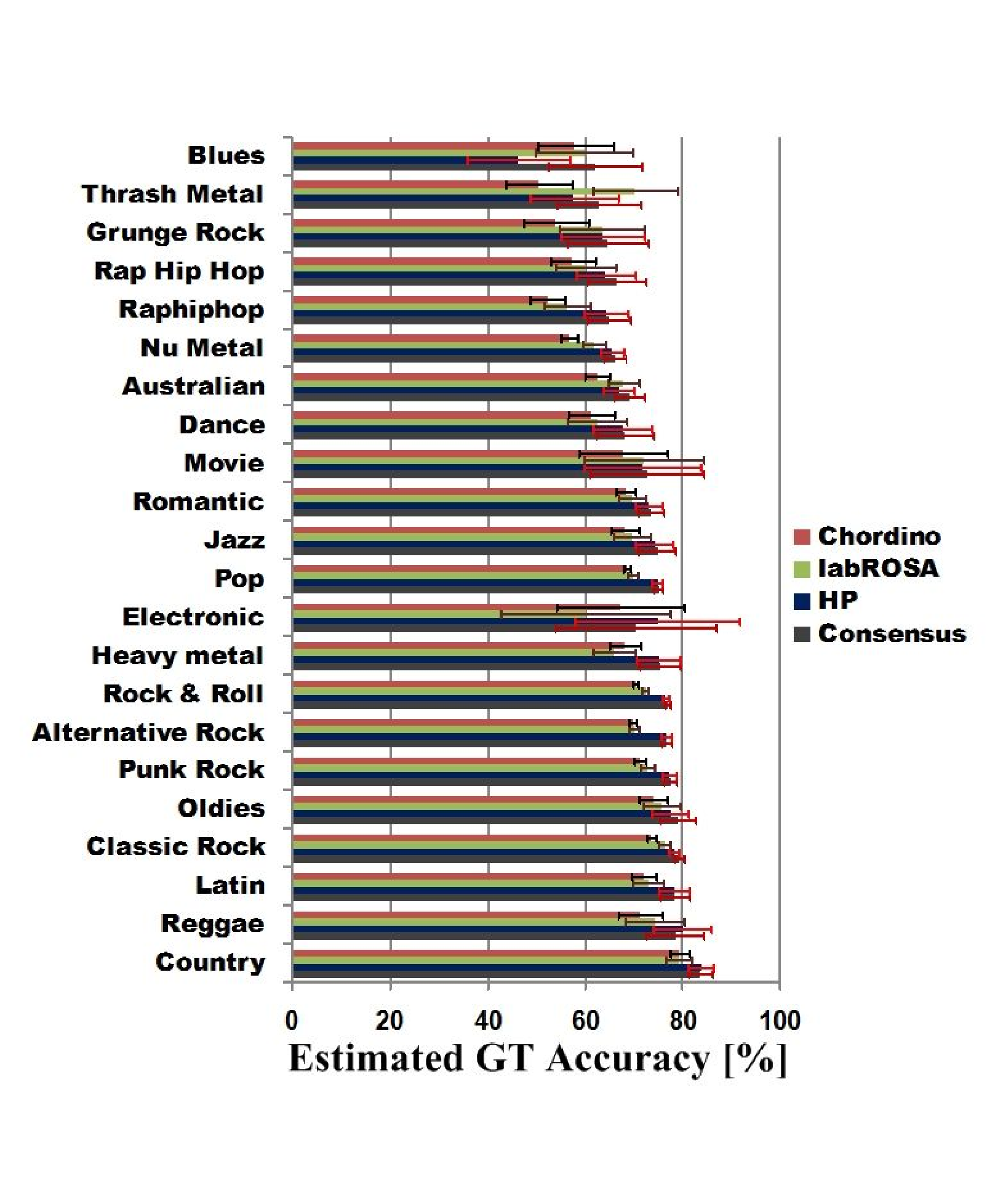}}
 \caption{Estimated GT accuracies of the test systems on each genre, using I model. }
 \label{fig:genre_performance_musicDB_individualGaussian}
\end{figure*}

\begin{figure*}
 \centerline{
 \includegraphics[scale=0.7]{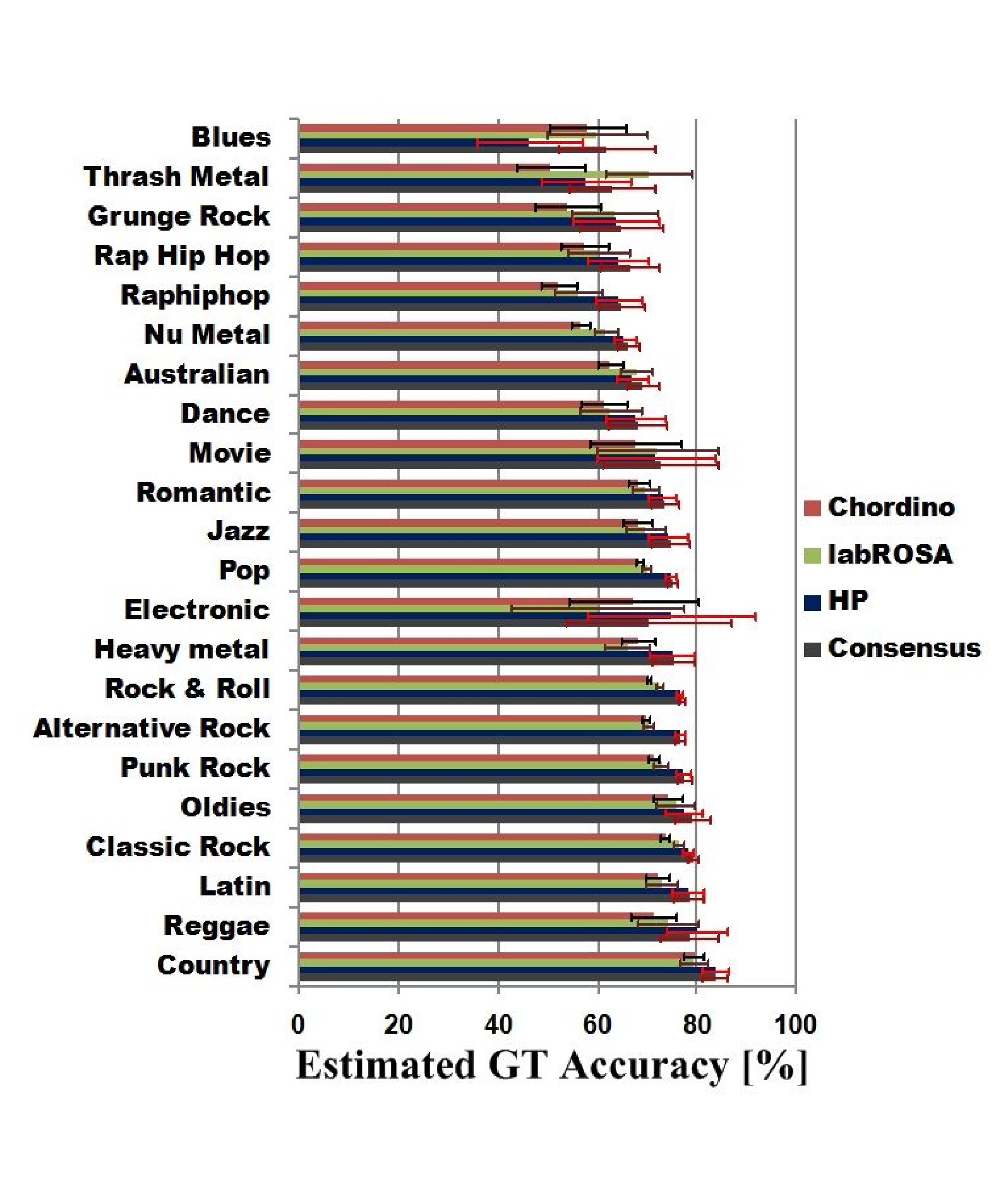}}
 \caption{Estimated GT accuracies of the test systems on each genre, using L model. }
 \label{fig:genre_performance_musicDB_linearRegression}
\end{figure*}

Finally, we categorized the test songs by their genre and estimated the mean GT accuracies of the test systems on each genre. The results are illustrated on Figure \ref{fig:genre_performance_musicDB_singleGaussian} to Figure \ref{fig:genre_performance_musicDB_linearRegression}. We observed that HP performs better on most of genres, especially on Rock related genres. This is consistent with the fact that the system is trained on songs mainly from the Rock genre. Meanwhile, the performances of labROSA and Chordino are highly overlapped, except for some genres containing few songs, which may happen by chance.

\subsection{Consensus}\label{sec:consensus}
Inspired by the fact that the best test system HP does not always outperform the other two in genre-specific estimation, we tried to combine predictions from the three systems so as to improve the recognition accuracy. As a trial, we simply combined predictions on each frame by majority vote, which yielded a consensus prediction of the three systems (denoted by ``Consensus"). The performance of the Consensus system are presented in Table \ref{tab:verification}, as well as Figure \ref{fig:estimated_GTacc_on_musicDB} to Figure \ref{fig:genre_performance_musicDB_linearRegression}. It is promising to see that the consensus prediction performs slightly better than HP (on I/L models), by means of compensating the low performance of HP on certain genres (e.g.~Funk and Blues). This observation is sufficiently encouraging that an investigation of combining systems' predictions using machine learning techniques will be carried out in the future.

\section{Conclusions and future work}
We have proposed a new method to evaluate chord recognition systems on songs which do not have full annotations. The approach goes beyond the existing evaluation metrics, allowing us to carry out extensive analysis on chord recognition systems, such as their generalizations to different genres. In the experiments, we tested this method on three systems, and the resulting confidence intervals on a validation set verified its reliability. We then evaluated these systems on a much larger test set and obtained some promising observations which can not be achieved by current evaluation techniques. These observations inspired us to combine predictions of different systems, and the resulting consensus system achieved the best performance by means of compensating weakness of a specific system.

For the future work, we aim at improving the reliability of the statistical models proposed. Since there may be errors and omissions in chord sequences obtained from the online databases, these chord sequences may become outliers in the validation and test sets (e.g.~circled points  in Figure \ref{fig:PACC_vs_GTACC}). A method to detect and remove these outliers is then a direction of our future work. Meanwhile, as pointed out in Section \ref{sec:consensus}, an investigation of combining systems' predictions using machine learning techniques will also be conducted.

\bibliography{ISMIR2011}

\end{document}